\documentclass[dvips]{article}

\usepackage{icrc2011,amsmath}

\title{Azimuthal modulation of cosmic ray flux as an effect of geomagnetic field\\
      in the ARGO-YBJ experiment}

\newcommand{\etal}{\MakeLowercase{\textit{et al. }}}
\shorttitle{P. Bernardini \etal Azimuthal modulation of cosmic rays}

\authors{P. Bernardini$^1$, A. D'Amone$^1$, H.H. He$^2$, 
        G. Mancarella$^1$, L. Perrone$^3$, A. Surdo$^4$,\\ 
	on behalf of the ARGO-YBJ Collaboration}
\afiliations{$^1$INFN and Dipartimento di Fisica dell'Universit\`a del Salento, Lecce, Italy\\
            $^2$Key Laboratory of Particle Astrophysics, Institute of High Energy Physics (CAS), Beijing, P.R. China\\
            $^3$INFN and Dipartimento di Ingegneria dell'Innovazione dell'Universit\`a del Salento, Lecce, Italy\\
            $^4$INFN, Lecce, Italy}
\email{paolo.bernardini@le.infn.it}

\abstract{The geomagnetic field causes not only the East-West effect on the primary 
cosmic rays but also affects the trajectories of the secondary charged particles in the
shower, causing their lateral distribution to be stretched along certain directions. 
Thus both the density of the secondaries near the shower axis and the trigger efficiency
of a detector array decrease. The effect depends on the age and on the direction of the
showers, thus involving the measured azimuthal distribution. Here the non-uniformity of
the azimuthal distribution of the reconstructed events in the ARGO-YBJ experiment is
deeply investigated for different zenith angles 
on the light of this effect. The influence of the geomagnetic field as well as 
geometric effects are studied by means of a 
Monte Carlo simulation.}

\keywords{ Magnetic Fields - Cosmic Rays - Extensive Air Showers }

\begin{document}
\maketitle


Cosmic rays (CRs) are charged particles and their paths are deflected by the magnetic fields.
The galactic magnetic field randomizes the CR arrival directions. The 
geomagnetic field restrains low-rigidity CRs from reaching the terrestrial atmosphere and
causes that the CR flux is lower from East than from West.
The geomagnetic field acts also on the charged particles of the extensive air showers (EAS)
during their path of few kilometers in the atmosphere. Cocconi~\cite{bib:cocco} suggested 
that the lateral displacement induced by the Earth magnetic field is not negligible with 
respect to the Coulomb scattering when the shower is young. Therefore the effect could
increase for high altitude observations. In particular the shower extension along the East-West 
direction 
is larger than along the North-South direction (the opposite for the particle density).
The different density of charged particles introduces an azimuthal modulation due to the 
different trigger efficiency of EAS detectors. This modulation was observed at the Yakutsk 
array for EAS with energy above $50\ PeV$~\cite{bib:ivan} and at the Alborz observatory 
for energy above $100\ TeV$~\cite{bib:iran}. A North-South asymmetry has been observed 
also in EAS radio-experiments~\cite{bib:coda} and it can be explained as an effect of the 
geomagnetic Lorentz force on the EAS charged particles. 
Therefore the geomagnetic influence on the lateral distribution must be taken into account
in the EAS simulation~\cite{bib:cillis} and it was also suggested that a pointing correction 
is necessary for \u{C}erenkov telescopes because of the geomagnetic field~\cite{bib:magic}.

The effect of the geomagnetic field on the trigger efficiency of the ARGO-YBJ experiment 
was studied and simulated~\cite{bib:hhh}. 
Due to the field $B$ the average shift ($d$) of an electron (or positron)
in the shower plane is
\begin{equation}
  d = \frac{q h^2 B sin \chi}{2 E_e cos^2 \theta}
  \label{eq:ddd}
\end{equation}
where $\chi$ is the angle between $\vec{B}$ and the particle velocity $\vec{v}$,
$q$ the electric charge, $h$ the average vertical height of the electron path, $E_e$ 
the average energy and $\theta$ the zenith angle. 
The trigger efficiency is connected to the size of the EAS footprint (lower particle
density corresponds to lower trigger efficiency). Therefore the stretching of the EAS lateral distribution
introduces a modulation with respect to the azimuth angle ($\phi$). The azimuth distribution 
is expected to be
\begin{equation}
\frac{dN}{d\phi}=N_0 \left\{1 + g_1 cos \left(          \phi+\phi_1         \right) + 
                                 g_2 cos \left[ 2 \left( \phi+\phi_2 \right) \right] \right\}
\label{eq:harmo2}
\end{equation}
with $\phi_1 = \phi_2 = - \phi_B$ ($\phi_B$ is the azimuth of the geomagnetic field), 
$g_1 \propto sin 2\theta$ and $g_2 \propto sin^2 \theta$ if the modulation is totally
geomagnetic. Furthermore~\cite{bib:hhh} 
\begin{equation}
\frac{g_1}{g_2} = \frac{sin 2\theta_B}{sin^2 \theta_B} \times \frac{sin 2\theta}{sin^2 \theta}
\label{eq:thetaB}
\end{equation}
where $\theta_B$ is the zenith angle of the geomagnetic field.

\section{ARGO-YBJ experiment}
ARGO-YBJ 
is a full-coverage EAS experiment, located close to the YangBaJing
village in Tibet (People's Republic of China) at $4300\ m$ above sea level. Its geographical
coordinates are $90^\circ 31' 50'' E$ and $30^\circ 06' 38'' N$. The experiment is mainly 
devoted to Very High Energy (VHE) $\gamma$-astronomy and CR studies. The detector is essentially 
a continuous carpet ($78 \times 74\ m^2$) of Resistive Plate Counters. The detection area is 
enlarged to $110 \times 100\ m^2$ by means of a partially equipped guard ring. The time-space
pattern allows a detailed reconstruction of showers induced by gamma and charged primaries.

In the ARGO-YBJ reference system the azimuth angle of EAS is defined with respect to the detector
axes in the anticlockwise direction ($\phi = 0^\circ$ for showers aligned with the x-axis and moving
towards the positive direction). Thus the azimuth angle of showers going towards the geographical North 
is $\phi_N=71.96^\circ \pm 0.02^\circ$. The quoted error is due to the measurement of the
orientation of the detector axes with respect to the geographical reference system.

\begin{figure}[!t]
  \centering
  \includegraphics[width=2.8in]{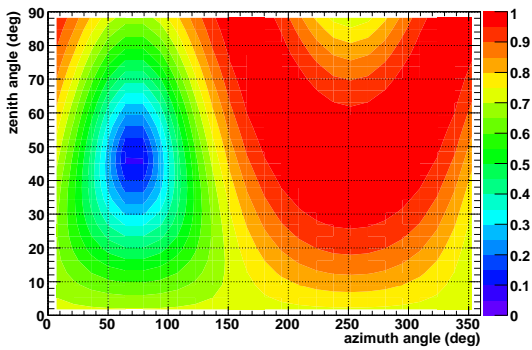}    
  \caption{Absolute value of $sin \chi$ versus
          local angular coordinates ($\theta$ and $\phi$) of the
          arrival direction of a charged particle in the ARGO-YBJ
          reference frame. $\chi$ is the angle between the magnetic
          field at YangBaJing and the velocity of the particle.} \label{fig:sinchi}
\end{figure}

According to the International Geomagnetic Reference Field (IGRF) model available on the NOAA web
site~\cite{www:noaa} at YangBaJing the geomagnetic field ($B=49.7\ \mu T$) has the following
angular coordinates in the ARGO-YBJ reference system
\begin{equation}
\theta_B = 46.4^\circ, \ \ \ \ \ \ \ \ \ \ \ \phi_B = 71.89^\circ.
\label{eq:angB}
\end{equation}
The geomagnetic effect on primary CRs is negligible for the EAS collected by ARGO-YBJ. Thus we 
focus on the effects on the secondary particles in the shower. The Lorentz force acting on
charged particles depends on the angle $\chi$ used in Eq.~(\ref{eq:ddd}). The
absolute value of $sin \chi$ is shown in Fig.~\ref{fig:sinchi} as a function of $\theta$ and 
$\phi$. It has been checked that the variation with the height is negligible. In fact 
from $4.3\ km$ up to 
$30\ km$ above sea level the product $B sin \chi$ changes less than $2 \%$.

\section{Data analysis}

This analysis is based on the data collected in 3 days (October 12-14, 2010) with the 
trigger condition that at least 20 time-pixels (pads) are fired. In order to get a 
reliable reconstruction of the shower direction, the following cuts have been applied:
shower core reconstructed inside a square of $40 \times 40\ m^2$ at the center of the
carpet, zenith angle lower than $60^\circ$. After these cuts $\sim 130$ millions of 
events are selected.

The timing calibration of the pads has been performed according to the Characteristic
Plane method~\cite{bib:cp1}. This procedure removes the systematic time differences
and arranges the time correction in order to make null the mean value of the direction
cosines. After this step a last premodulation correction is necessary in order to remove
a small over-correction intrinsic to the method~\cite{bib:cp2}. It is remarkable that
small systematics in the timing calibration prevent the azimuthal analysis and the 
premodulation correction is crucial in order to get the proper $\phi$-distribution.

\begin{figure}[!b]
  \centering
  \includegraphics[width=2.8in]{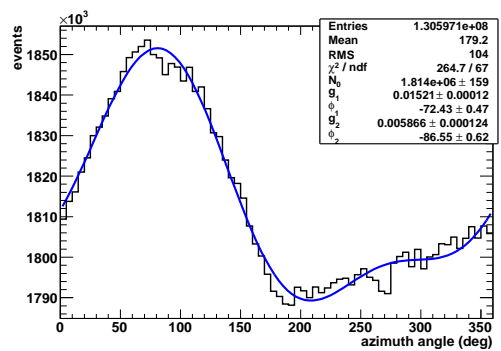}    
  \caption{Azimuthal distribution of EAS sample. 
          Fit with function~(\ref{eq:harmo2})
	  is superimposed.} \label{fig:harmo}
\end{figure}


The azimuthal distribution is shown in Fig.~\ref{fig:harmo} together with a fit performed
according to function~(\ref{eq:harmo2}). $\chi^2/ndf \sim 4$ and this is due to a small 
inefficiency at $\phi \sim n \times 90^\circ$ ($n=0,1,2,3,4$) which has not been examined here 
but does not invalidate this analysis. The fit results are 
$$ g_1 = (1.521 \pm 0.012) \%, \ \ \ \ \ \ \ \phi_1 = -72.43^\circ \pm 0.47^\circ \simeq - \phi_B, $$
$$ g_2 = (0.587 \pm 0.012) \%, \ \ \ \ \ \ \ \phi_2 = -86.55^\circ \pm 0.60^\circ \neq   - \phi_B. $$
The first coefficient is 3 times higher than the second one. The phase of the first harmonic 
results fully compatible with what expected. This is not the case for the phase of the second
harmonic.

\begin{figure}[!t]
  \centering
  \includegraphics[width=2.8in]{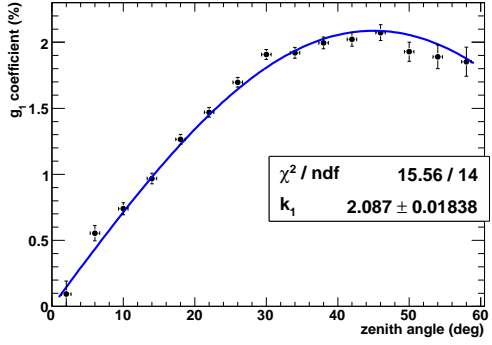}    
  \caption{Coefficient $g_1$ versus the zenith angle.
          Function $g_1 = k_1\ sin 2\theta$ 
	  is used for the superimposed fit.} \label{fig:triple}
\end{figure}

The geomagnetic origin of the modulation can be checked also by studying the dependence of the harmonic 
coefficients on the zenith angle. According to~\cite{bib:hhh} 
\begin{equation}
  g_1 = k_1\ sin  2\theta, \ \ \ \ \ \ \ g_2 = k_2\ sin^2 \theta. \label{eq:g12shape}
\end{equation}
This behaviour is verified for $g_1$ (see the fit in Fig.~\ref{fig:triple}), not for $g_2$
(the plot is missing for room saving).

The unexpected result for 
the second harmonic cannot be explained as a 
consequence of some systematic error. The disagreement can be solved simply assuming
a mix of geomagnetic and detector effects. The periodicity of a detector effect must
be $180^\circ$ and then acts only on the second harmonic, its phase must be $0^\circ$
or $\pm 90^\circ$ in connection with $x$ and $y$-axes. The fit of Fig.~\ref{fig:harmo}
suggests the second solution because $\phi_2 = -86.55^\circ$ is in-between $-\phi_B$
and $-90^\circ$.

\begin{figure*}[!t]
  \centerline{\includegraphics[width=2.4in]{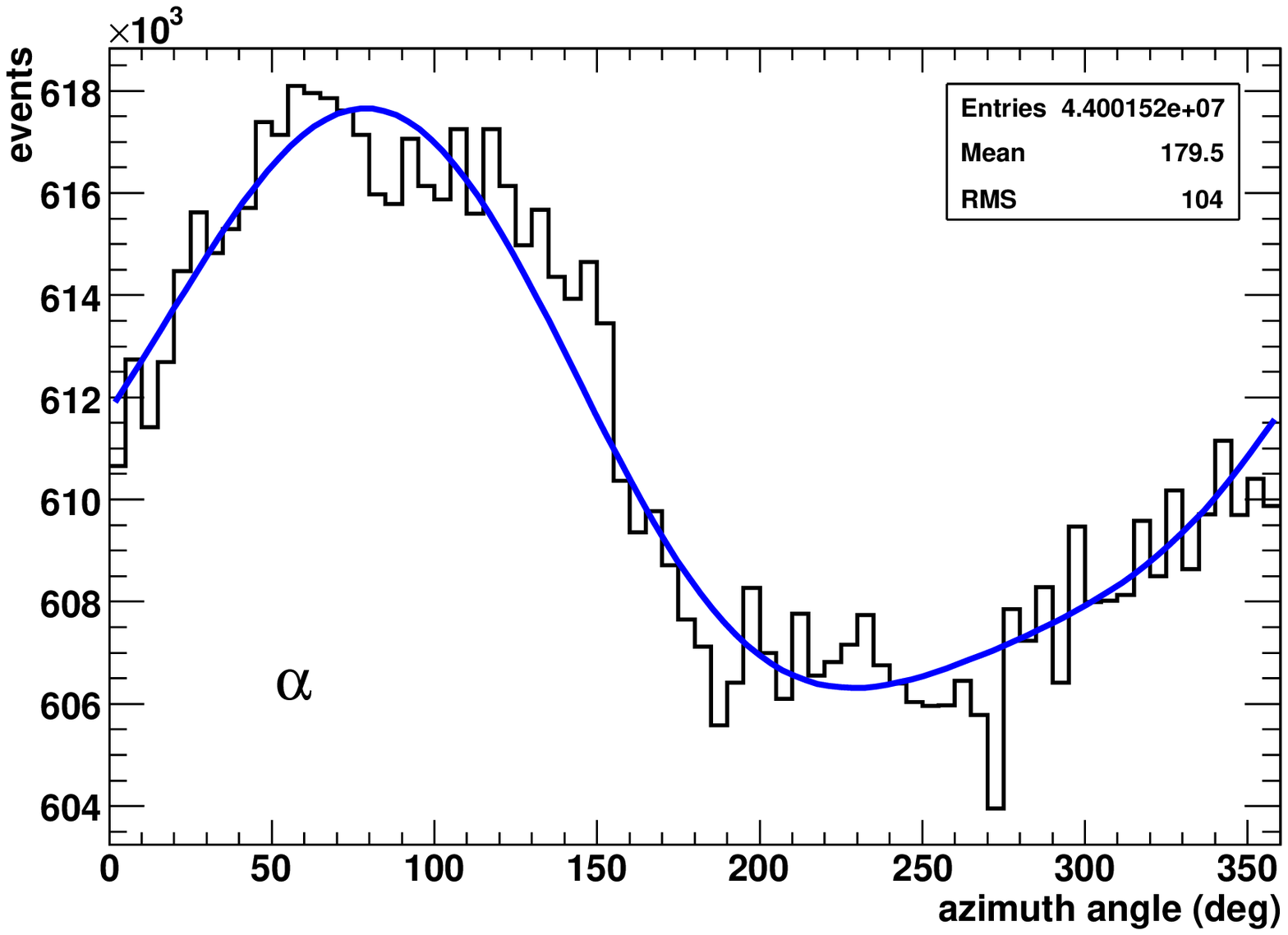}\label{fig:0020} \hfil
             \includegraphics[width=2.4in]{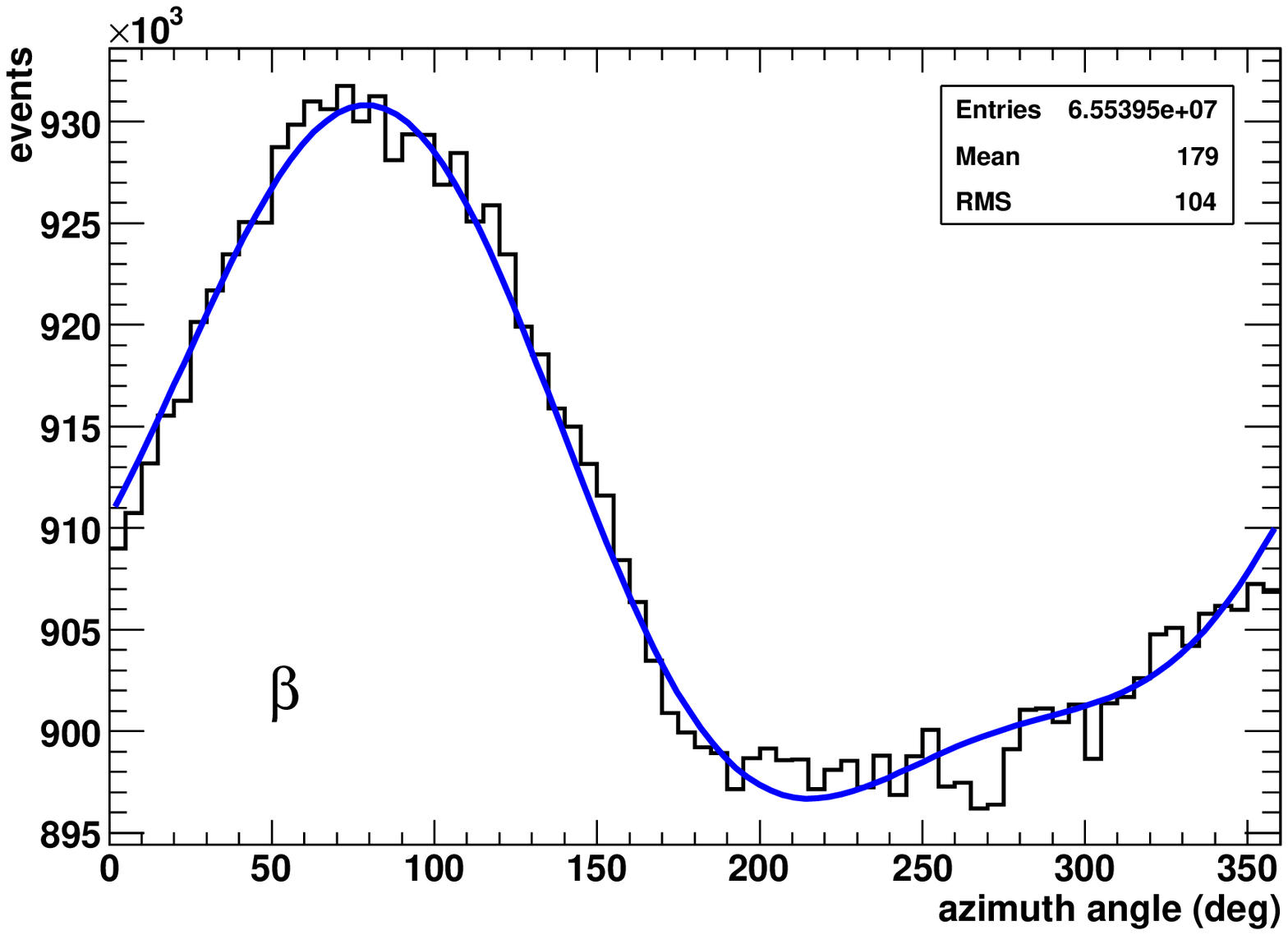}\label{fig:2040} \hfil
             \includegraphics[width=2.4in]{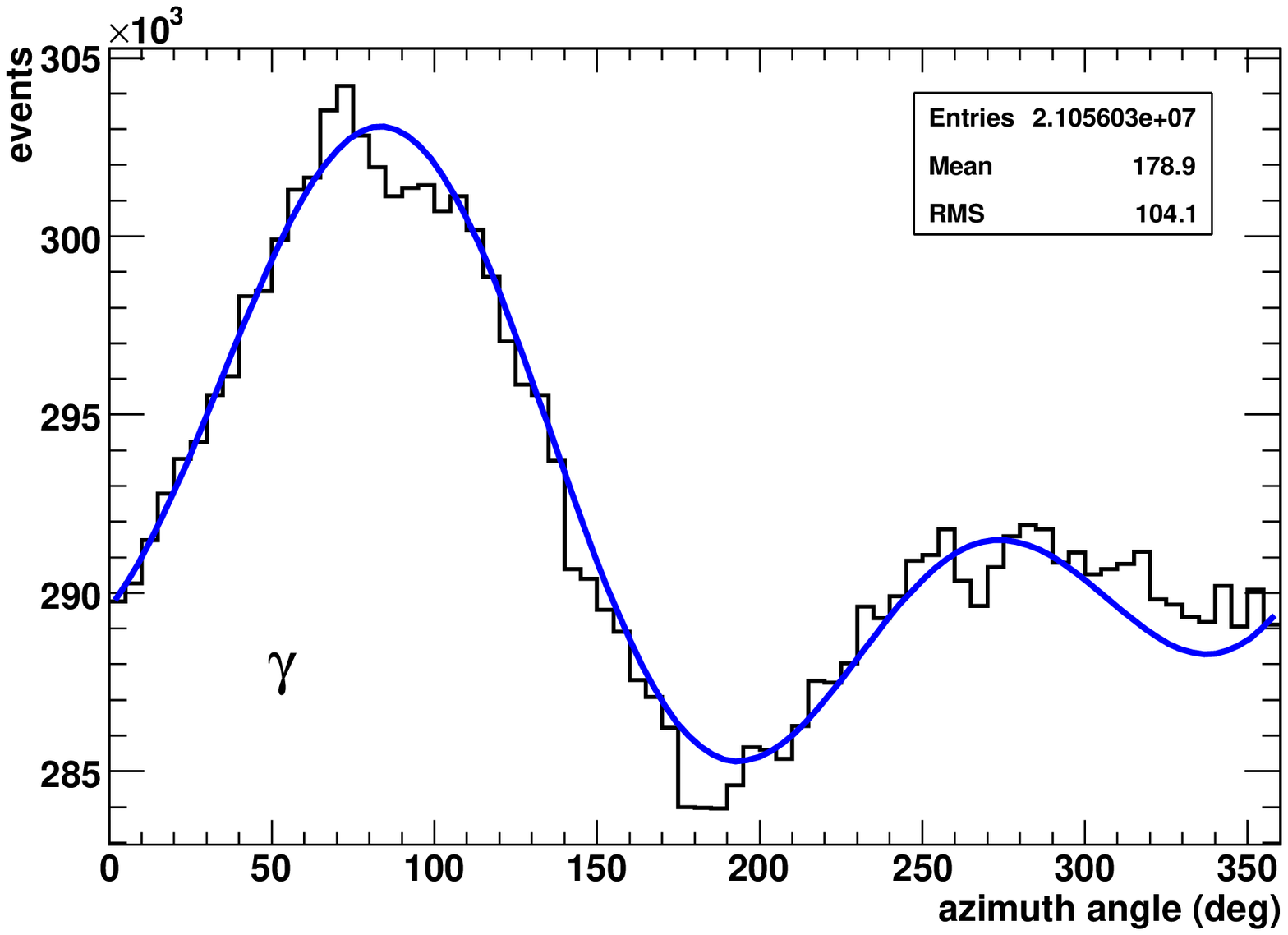}\label{fig:4060} }
  \caption{Azimuthal distribution in different $\theta$-ranges
          ($0^\circ-20^\circ$, $20^\circ-40^\circ$ and $40^\circ-60^\circ$).
	  Fit with function~(\ref{eq:harmo3}) is superimposed.} \label{fig:tripla}
\end{figure*}

In this hypothesis the second harmonic can be splitted in two parts: one ($2B$) is due 
to the magnetic field, the other one ($2A$) can be originated by a detector asymmetry.
Three different data sets have been selected on the basis of the zenith value in order 
to disentagle these two effects. 
The $\phi$-distributions of the 3 subsamples ($\alpha$, $\beta$ and $\gamma$ in 
Fig.~\ref{fig:tripla}) can be fitted all together with a single function:
\begin{align}
  \frac{dN}{d\phi} &= N_i \left\{ 1  + k_1    \langle sin  2\theta \rangle_i\ cos \left(          \phi+\phi_1    \right)         \right.\nonumber\\
                    &\qquad \left. {} + k_{2B} \langle sin^2 \theta \rangle_i\ cos \left[ 2 \left( \phi+\phi_1    \right) \right] \right.\nonumber\\
                    &\qquad \left. {} + g_{2A}^i\ cos \left[ 2 \left( \phi+\phi_{2A} \right) \right] \right\}
\label{eq:harmo3}
\end{align}
where the coefficients of the magnetic component are written using Eq.s~(\ref{eq:g12shape}), 
the magnetic phase is the same for first and second harmonic and the index $i=\alpha,\beta,\gamma$
indicates the subsamples.
The quantities $N_i$, $\langle sin 2\theta \rangle_i$ and $\langle sin^2 \theta \rangle_i$
have been estimated separately for each subsample. Then the fit parameters are $k_1$, $k_{2B}$, 
$\phi_1$, $g_{2A}^\alpha$, $g_{2A}^\beta$, $g_{2A}^\gamma$ and $\phi_{2A}$. The result of this 
new fit is reported in Tab.~\ref{tab:fit}. The value of $k_1$ is compatible with the fit in 
Fig.~\ref{fig:triple}. The ratio $k_1/k_{2B} = 2.41 \pm 0.90$ is compatible with 1.9, 
value calculated according to~(\ref{eq:thetaB}) and~(\ref{eq:angB}). The magnetic phase is in
perfect agreement with the expectation ($\phi_1 = - \phi_B$). It is remarkable that 
$g_{2A}^i$ increases with $\theta$ and $\phi_{2A}$ is compatible with $-90^\circ$ as expected 
for a detector effect.

\begin{table}[t]
\begin{center}
\begin{tabular}{rr} \hline
 $k_1$            (\%)  & $2.101 \pm 0.016$ \\
 $k_{2B}$         (\%)  & $0.87  \pm 0.32$ \\ 
 $\phi_1$   ($^\circ$)  & $-71.86 \pm 0.44$ \\ \hline
 $g_{2A}^\alpha$  (\%)  & $0.142 \pm 0.024$ \\
 $g_{2A}^\beta$   (\%)  & $0.350 \pm 0.061$ \\
 $g_{2A}^\gamma$  (\%)  & $1.25  \pm 0.13$ \\
 $\phi_{2A}$ ($^\circ$) & $-92.5 \pm  2.9$ \\ \hline
 $\chi^2/ndf$            & 491/209  \\ \hline
\end{tabular}
\caption{Results of the fit with function~(\ref{eq:harmo3})
         of the distributions in Fig.~\ref{fig:tripla}.}\label{tab:fit}
\end{center}
\end{table}

A possible explanation of the detector origin of the harmonic $2A$ is that the pads have a different 
density along the detector axes ($1.54\ pads/m$ along $x$-axis and $1.76\ pads/m$ along $y$-axis).
Taking into account the trigger requirement (at least 20 pads fired in a time-window of $420\ ns$)
the different density can explain the higher trigger efficiency for showers along the $y$-axis
($\phi=\pm 90^\circ$). The increasing of the effect with the zenith angle is the expected 
consequence of this hypothesis.

In order to check the results for the first harmonic we observe according to function~(\ref{eq:harmo3})
that the mean values of the direction cosines $\langle l \rangle$ and $\langle m \rangle$ of the showers
depend on $k_1$ and $\phi_1$ as in the following
$$ \langle l \rangle = + \frac{k_1}{2} cos \phi_1 \langle sin 2\theta\ sin \theta \rangle, $$
$$ \langle m \rangle = - \frac{k_1}{2} sin \phi_1 \langle sin 2\theta\ sin \theta \rangle. $$
In the analyzed sample the mean values are
$$ \langle l                         \rangle = (  12.12 \pm 0.29) \times 10^{-4} $$
$$ \langle m                         \rangle = (  35.56 \pm 0.30) \times 10^{-4} $$
$$ \langle sin\ 2\theta\ sin\ \theta \rangle = (3588.39 \pm 0.20) \times 10^{-4} $$
As a consequence
$$ k_1 = (2.094 \pm 0.016) \%, \ \ \ \ \ \ \ \phi_1 = -71.18^\circ \pm 0.57^\circ $$
These values are fully compatible with the the fit results in Tab.~\ref{tab:fit}.

\section{Simulation}\label{sec:simu}
{\bf Beams} - In order to check the effect of the magnetic field on the data collection and on the
EAS reconstruction, some "beams" of primary protons have been simulated with $\theta = 45^\circ$,
$E = 3\ TeV$ and interacting at $19\ km$ of height. The Corsika code~\cite{bib:corsika} has been
used to simulate the shower development and a GEANT-based code~\cite{bib:geant} to reproduce the
detector response. Nine different beams have been generated with 3 different values of the azimuth
angle ($\phi=71.5^\circ$ where $sin \chi \sim 0$, $\phi=251.5^\circ$ where $sin \chi$ is 
maximum and the intermediate angle $\phi=161.5^\circ$) and assuming 3 different values of the 
magnetic field (null, the effective magnetic field and twice the effective magnetic field).
Negative and positive EAS components have been studied separately. It has been verified that 
the geomagnetic field separates positive and negative cores of about $4\ m$ right on the East-West
direction. This stretching of the lateral distribution does not affect the combined reconstruction
(positive + negative component) but acts on the trigger efficiency, according to~\cite{bib:hhh}.
The results are summarized in Tab.~\ref{tab:beam} and confirm that the trigger efficiency decreases 
as the geomagnetic field or the $sin \chi$ value increase. 

\begin{table}[t]
\begin{center}
\begin{tabular}{rc|ccc} \hline
$\phi \ \ \ \ $ & $sin\chi$ & $B = 0\ \mu T$ & $49.7\ \mu T$ & $99.4\ \mu T$ \\ \hline
$ 71.5^\circ$   & 0.02       & $29.3 \%$          & $29.4 \%$          & $29.3 \%$          \\
$161.5^\circ$   & 0.87       & $29.6 \%$          & $28.4 \%$          & $26.0 \%$          \\
$251.5^\circ$   & 1.00       & $29.0 \%$          & $27.7 \%$          & $24.9 \%$          \\ \hline
\end{tabular}
\caption{Trigger efficiency for simulated "beams" of CRs (see Sec.~\ref{sec:simu}).
        Statistical error is $0.1 \%$.}\label{tab:beam}  
\end{center}
\end{table}

\begin{figure}[!b]
  \centering
  \includegraphics[width=2.8in]{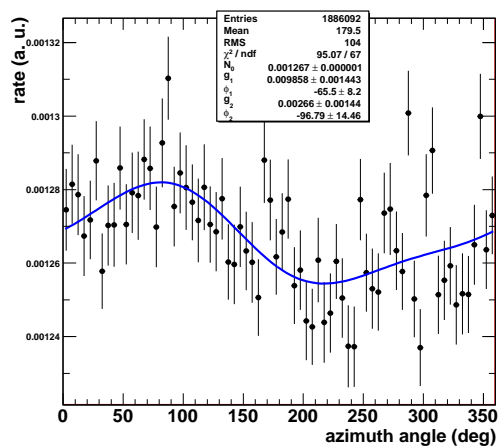}    
  \caption{Azimuthal distribution for simulated data
          and fit with function~(\ref{eq:harmo2}). The
          statistics does not allow to use function~(\ref{eq:harmo3}).} \label{fig:simu}
\end{figure}

{\bf Complete sample} - A sample of proton and helium-induced EAS has been simulated in the energy 
range $10\ GeV - 10\ PeV$, with a uniform azimuthal distribution and the proper geomagnetic field. 
The analysis chain of the real data has been applied to the simulated ones, with the exception of the
timing calibration that is not necessary for simulated data. The azimuthal distribution of triggered 
and selected events ($\sim 2 \times 10^6$) is not uniform (Fig.~\ref{fig:simu}). The shape is similar 
to that of Fig.~\ref{fig:harmo} and the parameters of the fit according to function~(\ref{eq:harmo2}) 
are almost compatible (the coefficients show larger discrepancies) but with large errors of the 
simulation fit. A larger simulation will be necessary in order to fix these few percent effects.

\section{Conclusions}
The modulation of the azimuthal distribution of a large EAS sample has been analyzed. 
The origin and the features of this modulation are fully understood. It is well described 
by means of two harmonics, the first one of the order of $1.5\%$, the second one of the 
order of $0.5\%$. The first harmonic is due to the geomagnetic Lorentz force on the shower 
charged particles. The second harmonic is the sum of magnetic and detector effects.

It is remarkable 
that in addition to the Moon shadow analysis the absolute pointing accuracy of EAS arrays could be tested
also with deep studies concerning the azimuthal modulation.


\begin{thebibliography}{}

\bibitem{bib:cocco} G. Cocconi, 
Phys. Review, 1954, {\bf 93}: 646-647. Erratum, Phys. Review, 1954, {\bf 95}: 1705-1706


\bibitem{bib:ivan} A.A. Ivanov et al., 
JETP Letters, 1999, {\bf 69}: 288-293

\bibitem{bib:iran} M. Bahmanabadi et al., Experim. Astronomy, 2002, {\bf 13}: 39-57.
M. Khakian Ghomi et al., 
Proceedings of 30th Intern. Cosmic Ray Conference, Merida, 2007 

\bibitem{bib:coda} D. Ardouin et al. (CODALEMA Collaboration),
Astropart. Physics, 2009, {\bf 31}: 192-200

\bibitem{bib:cillis} A. Cillis, S.J. Sciutto, 
J. Phys. G: Nucl. Part. Physics, 2000, {\bf 26}: 309-321

\bibitem{bib:magic} S.C. Commichau et al., 
Nuclear Instrum. Methods A, 2008, {\bf 595}: 572-586

\bibitem{bib:hhh} H.H. He et al., 
Proceedings of 29th Intern. Cosmic Ray Conference, Pune, 2005


\bibitem{www:noaa} www.ngdc.noaa.gov 

\bibitem{bib:cp1} H.H. He et al., 
Astropart. Physics, 2007, {\bf 27}: 528-532

\bibitem{bib:cp2} G. Aielli et al. (ARGO-YBJ Collaboration),
Astropart. Physics, 2009, {\bf 30}: 287-292

\bibitem{bib:corsika} www-ik.fzk.de/corsika/

\bibitem{bib:geant} wwwasd.web.cern.ch/wwwasd/geant/

\end{thebibliography}
\end{document}